\documentclass[journal]{IEEEtran}
\usepackage{cite}
\usepackage{xcolor}
\usepackage{graphicx}
\usepackage{amssymb,amsmath,bm}
\usepackage[colorlinks,linkcolor=black,anchorcolor=black,citecolor=black,urlcolor=black]{hyperref}
\usepackage{hyperref}
\usepackage{multirow}
\usepackage{amsfonts}
\usepackage{url}
\usepackage{multirow}
\usepackage{color}
\usepackage{tabularx}
\usepackage{color}
\usepackage{diagbox}
\usepackage{bbding}
\usepackage{comment}
\usepackage{enumitem}
\usepackage{float}
\ifCLASSINFOpdf
\else
\fi
\begin{document}
%
\title{Universal Discrete-Domain Speech Enhancement}
%
%
%

\author{Fei Liu, Yang~Ai,~\IEEEmembership{Member,~IEEE}, Ye-Xin Lu, Rui-Chen Zheng, Hui-Peng Du,~Zhen-Hua~Ling,~\IEEEmembership{Senior Member,~IEEE}
\thanks{This work was funded by the National Nature Science Foundation of China under Grant 62301521, the Anhui Provincial Natural Science Foundation under Grant 2308085QF200.}
\thanks{F. Liu, Y. Ai, Y.-X. Lu, R.-C. Zheng, H.-P. Du and Z.-H. Ling are with the National Engineering Research Center of Speech and Language Information Processing, University of Science and Technology of China, Hefei, 230027, China (e-mail: fliu215@mail.ustc.edu.cn, yangai@ustc.edu.cn, \{yxlu0102, zhengruichen, redmist\}@mail.ustc.edu.cn, zhling@ustc.edu.cn).}
\thanks{Corresponding author: Yang Ai.}
}
%
%

\markboth{}%
{Shell \MakeLowercase{\textit{et al.}}: Bare Demo of IEEEtran.cls for Journals}
%



\maketitle

\begin{abstract}
In real-world scenarios, speech signals are inevitably corrupted by various types of interference, making speech enhancement (SE) a critical task for robust speech processing. 
However, most existing SE methods only handle a limited range of distortions, such as additive noise, reverberation, or band limitation, while the study of SE under multiple simultaneous distortions remains limited. This gap affects the generalization and practical usability of SE methods in real-world environments.
To address this gap, this paper proposes a novel Universal Discrete-domain SE model called UDSE.
{UDSE primarily works to enhance speech by predicting clean discrete tokens that are quantized by the residual vector quantizer (RVQ) of a pre-trained neural speech codec and are predicted in accordance with the RVQ rules.}
Specifically, UDSE first extracts global features from the degraded speech. 
Guided by these global features, the clean token prediction for each VQ follows the rules of RVQ, where the prediction of each VQ relies on the results of the preceding ones. 
Finally, the predicted clean tokens from all VQs are decoded to reconstruct the clean speech waveform. 
During training, the UDSE model employs a teacher-forcing strategy, and is optimized with cross-entropy loss. 
Experimental results confirm that the proposed UDSE model can effectively enhance speech degraded by various conventional and unconventional distortions, e.g., additive noise, reverberation, band limitation, clipping, phase distortion, and compression distortion, as well as their combinations. 
These results demonstrate the superior universality and practicality of UDSE compared to advanced regression-based SE methods.

\end{abstract}


%
\IEEEpeerreviewmaketitle

\section{Introduction}

\IEEEPARstart{S}
{peech} signals inevitably suffer from various types of distortion in practical applications \cite{benesty2008microphone}. 
For instance, in outdoor environments, speech is frequently corrupted by different types of additive noise.
In transmission scenarios, when speech is transmitted between narrowband communication devices, only the low-frequency components are typically preserved, resulting in bandwidth limitation and the introduction of additional compression noise \cite{itu2001712}. 
These interferences significantly degrade the quality of the speech signal, posing challenges to the understanding of the target speech. 
Speech enhancement (SE) aims to process degraded speech signals using specific techniques to remove noise, reverberation, and other interfering components, with the goal of improving speech clarity and intelligibility \cite{loizou2007speech}.
The enhanced high-quality speech can better serve downstream tasks such as speech communication \cite{rabiner1995impact,mahdi2009advances,ince1992overview}, hearing aid devices, and automatic speech recognition (ASR) \cite{wang2019bridging,chen2015integration,delcroix2013speech}. 
High-quality, efficient, and generalizable SE techniques have become key research objectives for scholars worldwide, carrying significant academic and practical value.

In the past decade, with the development of deep learning, neural network-based SE models have significantly surpassed traditional algorithms in terms of enhanced speech quality and have gradually become the mainstream solutions for SE \cite{wang2018supervised}. 
At present, most neural SE models operate in the continuous domain, formulating SE as a regression task in which the neural network directly predicts clean speech waveforms or continuous spectral features and employs regression-based loss functions. 
Regression-based continuous-domain neural SE models \cite{tan2018convolutional,pascual2017segan,rethage2018wavenet,wang2021tstnn,luo2019conv,liu2022voicefixer,defossez2020real,serra2022universal,su2021hifi,ephraim2003speech,xu2014regression,kim2020t,yu2022dual,chu2021speech,abdulatif2024cmgan,lu2023mp,lu2025explicit} can be roughly categorized into waveform-modeling-based and spectrum-modeling-based approaches according to their prediction targets and modeling objectives. 
Waveform-modeling-based models directly predict clean speech waveforms from degraded ones using a single neural network \cite{tan2018convolutional,pascual2017segan,rethage2018wavenet,wang2021tstnn,luo2019conv,liu2022voicefixer,defossez2020real,serra2022universal}. 
For example, SEGAN \cite{pascual2017segan} is an early representative model that employs a fully convolutional network to directly predict clean speech waveforms and introduces a discriminator to enable adversarial training. 
DEMUCS \cite{defossez2020real} employs a multi-layer convolutional encoder-decoder architecture with U-Net-style skip connections, along with a sequence modeling module applied to the encoder output, to directly predict the clean speech waveform. 

However, direct prediction of clean speech waveforms often leads to low generation efficiency and high computational complexity. 
In contrast, spectrum-model-based models \cite{su2021hifi,ephraim2003speech,xu2014regression,kim2020t,yu2022dual,chu2021speech,abdulatif2024cmgan,lu2023mp,lu2025explicit} predict continuous clean spectral features and focus on enhancing speech at the frequency level, thereby avoiding direct manipulation of waveforms. 
Early spectrum-modeling-based SE models only enhance the amplitude spectrum while ignoring phase degradation, resulting in limited speech quality improvement \cite{xu2014regression,kim2020t,yu2022dual}. 
To further improve phase enhancement, recent studies have proposed directly enhancing the short-time complex spectrum, thereby enabling the implicit enhancement of both amplitude and phase components. 
For instance, CMGAN \cite{abdulatif2024cmgan} uses a Conformer-based \cite{gulati2020conformer} backbone to predict the amplitude spectrum and the real and imaginary parts of the short-time complex spectrum, which are then used to reconstruct the enhanced speech.
However, these methods still fall short in achieving explicit and accurate phase enhancement. 
To address this, in our previous work, we proposed MP-SENet \cite{lu2023mp,lu2025explicit}, a model that performs parallel and direct enhancement of both the amplitude and phase spectra based on anti-wrapping phase prediction \cite{ai2023neural,ai2024low}, demonstrating impressive results.

Recently, the rapid development of neural speech coding technologies has fostered innovation in the field of SE.
A few researchers have begun to explore the use of acoustic discrete tokens generated by neural speech codecs for SE, introducing novel discrete-domain approaches that aim to address classification problems rather than regression problems. 
In this framework, a neural network predicts the quantized acoustic discrete tokens (i.e., classification categories) of clean speech from degraded speech waveforms or continuous features.  
The enhanced speech is then reconstructed by decoding these predicted discrete tokens. 
The loss function is defined as the classification loss between the predicted and the ground truth clean discrete tokens.
However, these methods are still in the early stages of development and face numerous challenges.
For example, most existing work \cite{wang2024speechx,xue2024low,wang2024selm,yao2025gense,kang-etal-2025-llase} relies on the capabilities of large language models (LLMs) for token predictions, resulting in excessive model complexity that hinders the practical deployment of SE systems. 
Genhancer \cite{yang2024genhancer} employs a relatively simple DF-Conformer \cite{koizumi2021df} architecture but still relies on auxiliary components, such as ASR tools and self-supervised speech models, to extract text and semantic tokens for assisting acoustic token predictions, resulting in a complex and impractical solution. 

In the past one to two years, researchers have started to place greater emphasis on universal SE, aiming to improve the robustness and adaptability of SE models across various distortion types and complex scenarios.
For example, the URGENT Challenge \cite{zhang2024urgent}, which was launched in 2024 and has been held for two consecutive editions, aims to advance universal SE techniques targeting multiple distortion types under challenging real-world conditions.
{However, current regression-based continuous-domain SE models primarily focus on addressing conventional distortions, such as additive noise, reverberation, and band limitation \cite{kim2023hd,yang2024dm,nair2021cascaded}. 
These models often struggle to handle unconventional distortion types or scenarios involving multiple simultaneous interferences.}
{This is because many continuous-domain models formulate the task as a regression problem \cite{chu2021speech}, aiming to learn the underlying noise patterns and to fit an explicit input–output mapping through neural networks.}
{Such models tend to perform poorly when facing correlated noise (e.g., compression artifacts), where the noise shares strong characteristics with the underlying speech signal and cannot be explicitly modeled or fitted.}
{In contrast, classification-based discrete-domain approaches theoretically hold promise for achieving universal SE.} 
{These methods formulate the problem as a classification task that maps the input into a latent discrete space. They leverage the strong generative capability of generative models to learn data distributions and then decode waveforms from the classified tokens.
Furthermore, the cross-entropy loss used in the discrete domain penalizes incorrect predictions more severely, which helps the model better distinguish between correct and incorrect outputs.}
{This theoretically makes them more flexible in handling diverse and complex distortions.}
Nonetheless, current research on discrete-domain methods remains largely focused on conventional additive noise, with limited exploration of their universality across a wider range of distortions. 

Motivated by the aforementioned challenges and prior works, we propose a novel Universal Discrete-domain SE (UDSE) model guided by a pre-trained neural speech codec with a residual vector quantizer (RVQ). 
The UDSE model treats SE as a discrete-domain classification task, focusing on the prediction of clean acoustic discrete tokens quantized by the neural speech codec without requiring additional guidance from text or semantic tokens, and eliminating the need for LLMs. 
Specifically, starting from a randomly initialized sequence of discrete tokens, UDSE predicts clean acoustic discrete tokens for each VQ in sequence according to the rules of RVQ, where the prediction of each VQ depends on the outputs of the preceding ones. 
This process is guided by global features extracted from the degraded speech as a conditioning mechanism. 
Finally, the predicted tokens from all VQs are decoded to reconstruct the enhanced speech waveform. 
In our experiments, we simulated six types of distortions, i.e., noise, reverberation, band-limiting, clipping, phase distortion, and compression distortion, as well as three mixed distortion scenarios. 
Both objective and subjective experimental results confirm that our proposed UDSE exhibits robust performance across nine distortion scenarios, consistently restoring high-quality clean speech. 
This demonstrates its superior universality and practicality compared to advanced continuous-domain SE models.

The main contributions of this work are as follows:
\begin{itemize}[leftmargin=*]
\item {The proposed UDSE model introduces a novel token prediction paradigm that aligns with the RVQ scheme and may also provide inspiration for other discrete-domain speech generation tasks.}
\item The proposed UDSE model is capable of handling a wider range of distortion types, demonstrating superior universality and promoting the practicality of SE methods in complex real-world scenarios.
\end{itemize}

This paper is organized as follows.
In Section \ref{sec:related}, we briefly review the advanced neural speech codecs and their applications in {discrete-domain} speech generation methods.
In Section \ref{sec:propose}, we provide the details of our proposed UDSE model. 
In Section \ref{sec:exp}, we present our experimental results.
Finally, we give conclusions in Section \ref{sec:con}.


\section{Related Work}
\label{sec:related}

The proposed UDSE integrates speech coding techniques and operates within the discrete domain.
Therefore, this section primarily reviews two relevant areas: neural speech coding, as UDSE leverages neural speech codecs for discrete token representations, and discrete-domain speech generation methods, since UDSE also falls under the broader category of discrete-domain speech generation.

\subsection{Neural Speech Coding}

{With the rise of deep learning, end-to-end neural speech codecs now markedly outperform traditional methods. They learn adaptive, compact representations that retain essential information and remove redundancy even at low bitrates, making them the dominant direction in speech coding.}
A central component of neural speech codecs is the learnable quantizer \cite{zeghidour2021soundstream,defossezhigh,kumar2023high,ai2024apcodec}.
High quantization quality, reflected by the close alignment between the encoder features and their quantized counterparts, enables the decoder to generate higher-quality reconstructed speech.
In early work, researchers proposed a vector quantization (VQ) method that utilizes a learnable codebook, where each feature vector is assigned to the codeword with the minimum Euclidean distance, allowing effective representation learning within a differentiable quantization process \cite{garbacea2019low}.
{Although this alleviates gradient-continuity issues, it often converges unstably and yields higher distortion.}
To address these limitations, RVQ was introduced and successfully applied in building the neural speech codec SoundStream \cite{zeghidour2021soundstream}. 
RVQ connects multiple VQ modules in a residual manner by iteratively computing and quantizing the residual errors from previous quantization steps. 
Compared to single-VQ methods, RVQ further reduces overall quantization loss through this residual quantization process.

Building on SoundStream \cite{zeghidour2021soundstream}, which is the predecessor of neural speech codecs with RVQ, researchers have introduced further improvements in model architecture and training strategies, leading to the development of impressive codecs such as EnCodec \cite{defossezhigh} and DAC \cite{kumar2023high}.
Recently, we have also proposed improved strategies regarding the coding targets, motivated by the observation that directly coding speech waveforms requires multiple downsampling and upsampling steps, leading to high computational complexity. 
For example, we introduced APCodec \cite{ai2024apcodec}, which encodes and decodes the amplitude and phase spectra of speech, and further proposed MDCTCodec \cite{jiang2024mdctcodec}, which regards the MDCT spectrum of speech as coding target. 
By shifting the coding target from the waveform level to the spectral level, these approaches significantly reduce computational complexity while maintaining high reconstructed speech quality.

\subsection{Universal Speech Enhancement}
{In real-world scenarios, distortion is highly diverse, and SE models designed for a single distortion type have limited effectiveness in practice. 
Therefore, exploring universal SE methods is of great importance. 
Recently, several generative SE approaches based on diffusion models have demonstrated strong performance \cite{lu2022conditional,lemercier2023storm,richter2023speech,universe,universe++}. For example, UniverSE++ \cite{universe++} combines score-based diffusion and adversarial training: the score-based diffusion model uses stochastic differential equations to gradually transform clean speech into noise, while in the reverse process, it learns to recover the clean speech. Adversarial training further helps the model generate more natural-sounding speech, achieving excellent performance across various distortion types.}


{In addition to the abovementioned diffusion-based SE methods, recent studies have started to focus on applying neural speech codecs to the SE field, constructing novel discrete-domain SE approaches.
This type of method has the potential to achieve universal SE by equivalently converting enhancement tasks of any distortion type into a classification problem over discrete representations.} 
The discrete-domain SE leverages the discrete tokens produced by neural speech codecs as a bridge between degraded and clean speech, enabling SE within a discrete representation space. 
For example, Genhancer \cite{yang2024genhancer} adopts a DF-Conformer as its core architecture and combines discrete tokens with a self-supervised speech model to perform SE in the discrete domain.
GenSE \cite{yao2025gense} employs a hierarchical modeling framework. 
It first uses an LLM to refine the semantic tokens extracted from degraded speech, producing enhanced semantic representations. 
These refined semantic tokens, along with the degraded semantic tokens and acoustic tokens of the degraded speech, are then jointly processed by another LLM to generate enhanced acoustic tokens, which are finally decoded to produce the enhanced speech.
{However, existing approaches often depend on large-scale models and complicated processing pipelines, which undermines their practical applicability, and there has also been insufficient exploration of their effectiveness for universal SE.}

\begin{figure*}
    \centering
    \includegraphics[width=1\linewidth]{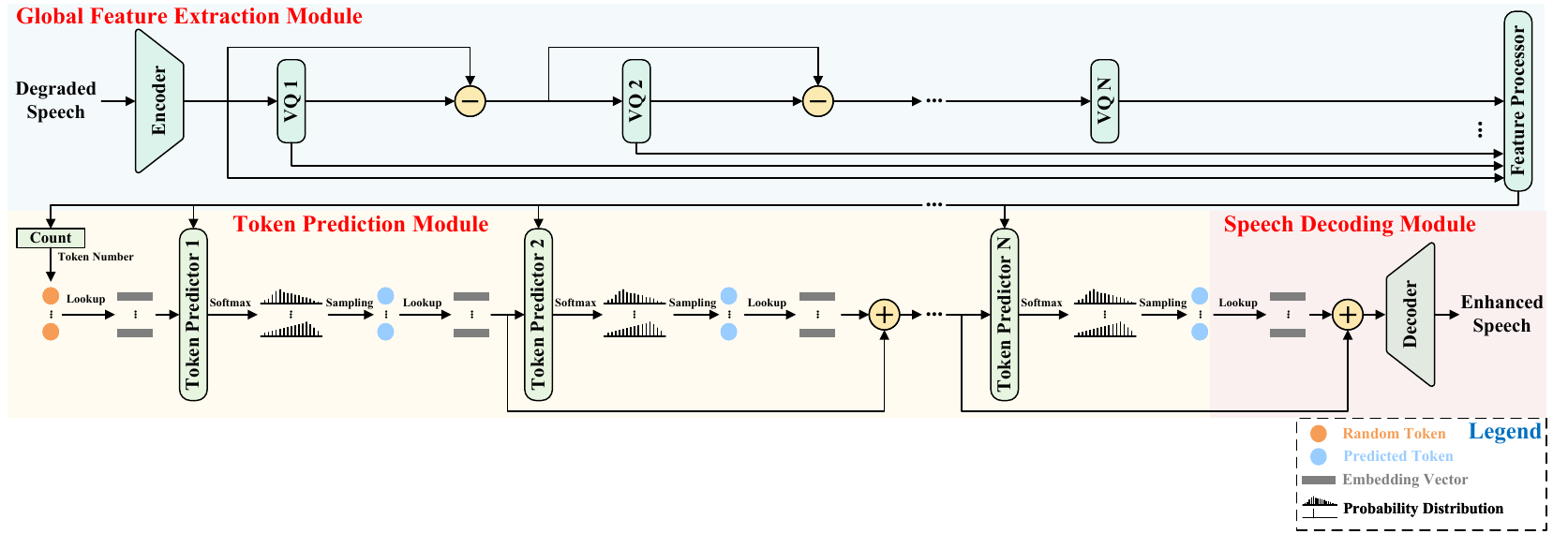}
    \caption{{Overview of the proposed UDSE's inference process.}}
    \label{fig:model}
\end{figure*}

\section{Proposed Method}
\label{sec:propose}
\subsection{Overview}

Suppose a clean speech $\bm{x}\in\mathbb R^T$ is affected by distortions $\theta\subset\Theta$, resulting in the degraded speech $\bm{y}\in\mathbb R^T$, i.e., 
\begin{equation}
    \bm{y}=\theta(\bm{x}),
\end{equation}
where $T$ is the length of speech waveform. 
$\Theta$ is the set of all possible distortions, and $\theta$ is a subset of $\Theta$. 
The SE aims to recover the clean speech $\hat{\bm{x}}\in\mathbb R^T$ from $\bm{y}$ by constructing enhancement method $\xi$ to minimize the gap between $\hat{\bm{x}}$ and $\bm{x}$, i.e.,
\begin{equation}
    \bm{\hat{x}}=\xi(\bm{y}).
\end{equation}
The ideal solution for universal SE is that $\xi$ is unique for all $\theta\subset\Theta$. 
The proposed UDSE model is designed to address the challenge of universal SE. 

\begin{figure*}
    \centering
    \includegraphics[width=1\linewidth]{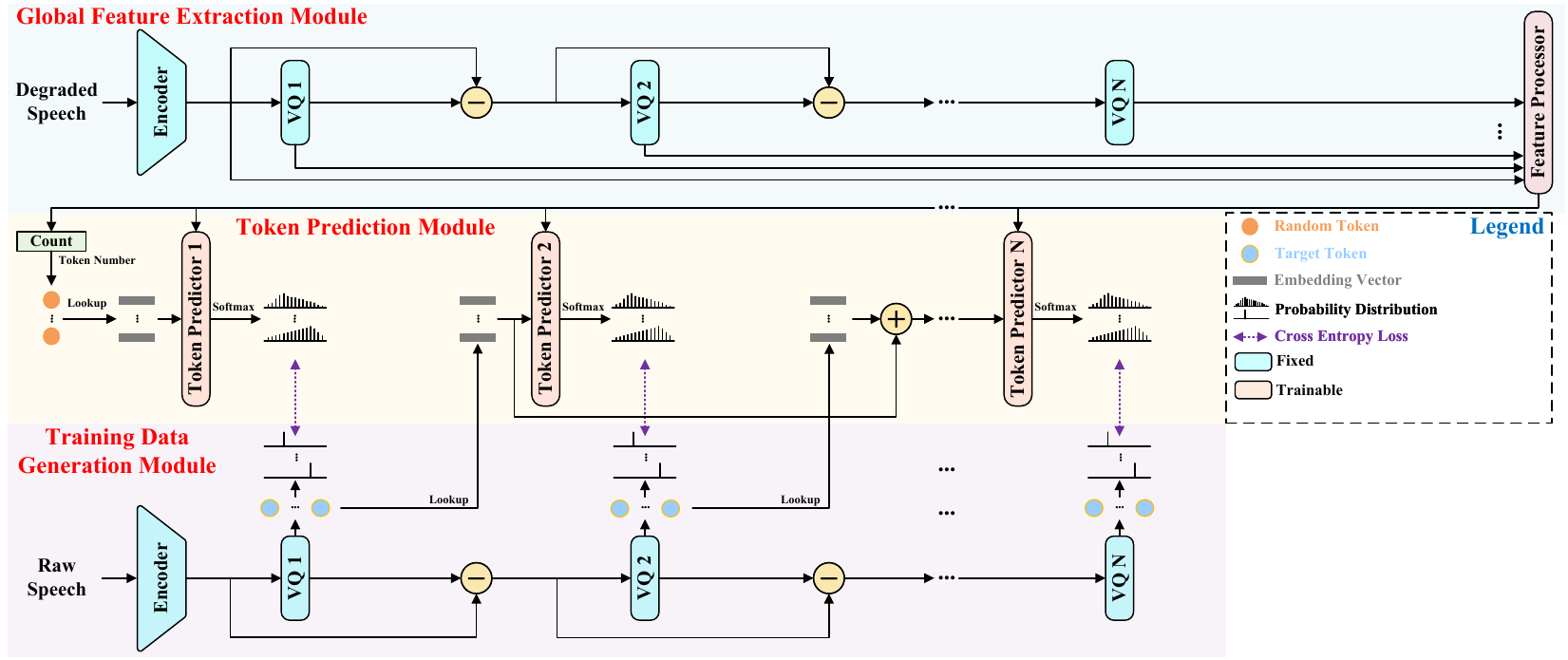}
    \caption{{Overview of the proposed UDSE's training process.}}
    \label{fig:train}
\end{figure*}

The proposed UDSE achieves SE in the discrete domain by solving a classification problem using a pre-trained neural speech codec with RVQ. 
As shown in Figure \ref{fig:model}, UDSE consists of multiple modules. 
The global feature extraction module extracts global features from $\bm{y}$ to guide the prediction of each VQ token in the token prediction module, which are finally decoded by the speech decoding module to generate $\bm{\hat{x}}$. 
As shown in Figure~\ref{fig:train}, the training phase of UDSE additionally incorporates a training data generation module, which provides input data and defines the classification loss.

\subsection{Global Feature Extraction Module}

The global feature extraction module is designed to extract global feature $\bm{G}^{(y)}\in\mathbb R^{C\times L}$ from the degraded speech $\bm{y}$, where $C$ and $L$ represent the global feature dimension and the number of frames, respectively. 
This module consists of an encoder $\phi_{E}$ and an RVQ with $N$ VQs $\phi_{Q_1},\dots,\phi_{Q_N}$ (each with the same codebook size) of a neural speech codec $\phi_{Codec}$, as well as a feature processor $\phi_{FP}$. 

Specifically, $\bm{y}$ is first passed through $\phi_{E}$ to obtain a downsampled encoded feature by a factor of $T/L$, i.e., 
\begin{equation}
\bm{E}^{(y)}=\phi_{E}(\bm{y}), 
\end{equation}
where $\bm{E}^{(y)}\in\mathbb R^{K\times L}$ and $K$ denotes the encoded feature dimension. 
Next, $\bm{E}^{(y)}$ is quantized by RVQ, and the quantization results of each VQ, $\bm{\tilde{Q}}^{(y)}_1,\dots,\bm{\tilde{Q}}^{(y)}_N\in\mathbb R^{K\times L}$, are sequentially outputted. 
During the quantization process, after the first VQ quantization, the difference between its input and output was used as the input to the second VQ. In other words, each subsequent VQ quantized the residual error of the previous one, forming a progressively refined quantization process.
For $\phi_{Q_1}$, its input is the encoded feature $\bm{E}^{(y)}$, let $\bm{Q}^{(y)}_1=\bm{E}^{(y)}$, i.e., 
\begin{equation}
\bm{\tilde{Q}}^{(y)}_{1}=\phi_{Q_1}(\bm{Q}^{(y)}_1),
\end{equation}
whereas for $\phi_{Q_n}$, its input is the quantization residual of $\phi_{Q_{n-1}}$, where $n=2,\dots,N$, i.e.,
\begin{gather}
\bm{Q}^{(y)}_n=\bm{Q}^{(y)}_{n-1}-\bm{\tilde{Q}}^{(y)}_{n-1},\\
\bm{\tilde{Q}}^{(y)}_{n}=\phi_{Q_n}(\bm{Q}^{(y)}_{n}).
\end{gather}
Finally, to fully preserve the original information, the quantization outputs from each VQ along with $\bm{E}^{(y)}$ are fed into the feature processor $\phi_{FP}$.  
Within $\phi_{FP}$, the input features are concatenated, dimension-reduced, and passed through $B_G$ Conformer blocks \cite{gulati2020conformer}, with residual connections applied to each block to link its input and output. 
This process generates the global feature $\bm{G}^{(y)}$, i.e,
\begin{equation}
    \bm{G}^{(y)}=\phi_{FP}(\bm{E}^{(y)},\bm{\tilde{Q}}^{(y)}_{1},\dots,\bm{\tilde{Q}}^{(y)}_{N}).
\end{equation}

\subsection{Token Prediction Module} 
The token prediction module starts from a randomly initialized sequence of discrete tokens, i.e.,
\begin{equation}
    \bm{d}_0=[d_{0,1},\dots,d_{0,L}]^\top, 
\end{equation}
and conditioned on the global feature $\bm{G}^{(y)}$, sequentially predicts the clean speech's token sequences generated through $\phi_{Q_1},\dots,\phi_{Q_N}$ by quantizing $\bm{x}$, i.e.,
\begin{equation}
    \hat{\bm{d}}_n^{(x)}=[\hat d_{n,1}^{(x)},\dots,\hat d_{n,L}^{(x)}]^\top, n=1,\dots,N,
\end{equation}
where $d_{0,*}/\hat d_{*,*}^{(x)}\in\{1,2,\dots,M\}$.
The token number $L$ is determined by the number of frames in $\bm{G}^{(y)}$ and $M$ is the codebook size of a VQ. 
The token prediction module consists of $N$ token predictors $\phi_{TP_1},\dots,\phi_{TP_N}$, each of which has a backbone composed of $B_T$ Conformer blocks \cite{gulati2020conformer} with residual connections.

Specifically, for the first token predictor $\phi_{TP_1}$, its input is the result $\bm{\hat{Q}}_{0}^{(x)}\in\mathbb{R}^{K\times L}$ obtained by looking up a random codebook $\mathbb W_0=\{\bm{w}_{0,m}\in\mathbb R^K|m=1,\dots,M\}$ using the random token $\bm{d}_0$, and the global feature $\bm{G}^{(y)}$. 
For subsequent token predictors $\phi_{TP_n}$ ($n=2,\dots,N$), following the dependency relationships among VQs in RVQ, their input is the sum of the results $\bm{\hat{Q}}_{1}^{(x)},\dots,\bm{\hat{Q}}_{n-1}^{(x)}$, which are obtained by looking up the codebooks $\mathbb W_1,\dots,\mathbb W_{n-1}$ of $\phi_{Q_1},\dots,\phi_{Q_{n-1}}$ using the previously predicted tokens $\bm{\hat{d}}_{1}^{(x)},\dots,\bm{\hat{d}}_{n-1}^{(x)}$, respectively, and the global feature $\bm{G}^{(y)}$. 
This also clearly reflected that our prediction was based on the RVQ principle.
Therefore, the output $\bm{\hat{U}}^{(x)}_{n}=[\bm{\hat{u}}^{(x)}_{n,1},\dots,\bm{\hat{u}}^{(x)}_{n,L}]\in\mathbb{R}^{M\times L}$ of $\phi_{TP_n}$ ($n=1,\dots,N$) can be expressed as
\begin{align}
\label{a}
    \bm{\hat{U}}^{(x)}_{n}=\begin{cases}
        \phi_{TP_n}(\bm{\hat{Q}}_{n-1}^{(x)},\bm{G}^{(y)}),&n=1, \\
        \phi_{TP_n}(\sum_{n_{0}=1}^{n-1} \bm{\hat{Q}}_{n_{0}}^{(x)},\bm{G}^{(y)}),&n=2,\dots,N.
    \end{cases}
\end{align}
Next, each frame of $\bm{\hat{U}}^{(x)}_{n}$ is passed through a softmax layer to compute the classification probability distribution. 
Taking the $l$-th ($l=1,\dots,L$) frame as an example, we can get
\begin{equation}
\label{p}
\bm{\hat{p}}^{(x)}_{n,l}=softmax(\bm{\hat{u}}^{(x)}_{n,l}),
\end{equation}
where $\bm{\hat{p}}^{(x)}_{n,l}\in\mathbb{R}^{M}$. 
The token is sampled from $\bm{\hat{p}}^{(x)}_{n,l}$ based on the maximum probability, selecting the most likely category from the $M$ categories, i.e.,
\begin{equation}
\hat{d}^{(x)}_{n,l}=argmax(\bm{\hat{p}}^{(x)}_{n,l}).
\end{equation}
The above process is executed from $n=1$ to $n=N$ to achieve the prediction of discrete tokens $\bm{\hat{d}}_{1}^{(x)},\dots,\bm{\hat{d}}_{N}^{(x)}$.

\subsection{Speech Decoding Module}

The speech decoding module uses the decoder $\phi_D$ of the neural speech codec $\phi_{Codec}$ to convert token prediction results $\bm{\hat{d}}_{1}^{(x)},\dots,\bm{\hat{d}}_{N}^{(x)}$ into the enhanced speech $\hat{\bm{x}}$. 
Specifically, $\bm{\hat{d}}_{1}^{(x)},\dots,\bm{\hat{d}}_{N}^{(x)}$ first retrieve $\bm{\hat{Q}}_{1}^{(x)},\dots,\bm{\hat{Q}}_{N}^{(x)}$ from the codebooks $\mathbb W_1,\dots,\mathbb W_{N}$ of $\phi_{Q_1},\dots,\phi_{Q_{N}}$, respectively, and then add them together as the input to $\phi_D$, which outputs $\hat{\bm{x}}$, i.e.,
\begin{equation}
    \bm{\hat{x}}=\phi_{D}(\sum\nolimits_{n=1}^N \bm{\hat{Q}}_{n}^{(x)}).\\
\end{equation}

\subsection{Training Strategy}

During the training phase, we first train the neural speech codec $\phi_{Codec}$ which includes the encoder $\phi_E$, quantizers $\phi_{Q_1},\dots,\phi_{Q_N}$, and decoder $\phi_D$. 
As shown in Figure \ref{fig:train}, we then froze all the parameters of the codec $\phi_{Codec}$ and proceeded to train the remaining components, i.e., $\phi_{FP}$ and $\phi_{TP_1},\dots,\phi_{TP_N}$.
Compared to the inference process, UDSE introduces a training data generation module to provide training target and jointly train $\phi_{FP}$ and $\phi_{TP_1},\dots,\phi_{TP_N}$ at the training phase. 
To reduce the impact of error propagation on model training and enable the model to learn the correct patterns more quickly, we adopt a teacher forcing strategy during training. 
In other words, each token predictor takes the ground-truth clean token as input rather than the predicted one.

Specifically, in the training data generation module, clean speech $\bm{x}$ is processed through $\phi_E$ and $\phi_{Q_1},\dots,\phi_{Q_N}$ to generate ground-truth clean tokens, i.e.,
\begin{equation}
    \bm{d}_n^{(x)}=[d_{n,1}^{(x)},\dots,d_{n,L}^{(x)}]^\top, n=1,\dots,N,
\end{equation}
for both training targets and token predictors' input, where $d_{*,*}^{(x)}\in\{1,2,\dots,M\}$. 
Then, $\bm{d}_1^{(x)},\dots,\bm{d}_{N-1}^{(x)}$ lookup codebooks $\mathbb W_1,\dots,\mathbb W_{N-1}$ to obtain quantized results $\bm{\tilde{Q}}_{1}^{(x)},\dots,\bm{\tilde{Q}}_{N-1}^{(x)}$ as inputs to $\phi_{TP_2},\dots,\phi_{TP_N}$ for teacher forcing training, and the output is calculated according to Equation (\ref{a}), i.e.,
\begin{align}
    \bm{\tilde{U}}^{(x)}_{n}=\begin{cases}
        \phi_{TP_n}(\bm{\hat{Q}}_{n-1}^{(x)},\bm{G}^{(y)}),&n=1, \\
        \phi_{TP_n}(\sum_{n_{0}=1}^{n-1} \bm{\tilde{Q}}_{n_{0}}^{(x)},\bm{G}^{(y)}),&n=2,\dots,N.
    \end{cases} 
\end{align}
Next, each frame of $\bm{\tilde{U}}^{(x)}_{n}$ is passed through a softmax layer to compute the classification probability distribution. 
Taking the $l$-th ($l=1,\dots,L$) frame as an example, according to Equation (\ref{p}), we can get
\begin{equation}
\bm{\tilde{p}}^{(x)}_{n,l}=softmax(\bm{\tilde{u}}^{(x)}_{n,l}),
\end{equation}
where $\bm{\tilde{u}}^{(x)}_{n,l}$ is the $l$-th column vector in $\bm{\tilde{U}}^{(x)}_{n}$. 
Finally, token $d_{n,l}^{(x)}$ is one-hot encoded to generate the target probability distribution $\bm{p}_{n,l}^{(x)}$, and the cross-entropy loss is defined with respect to $\bm{\tilde{p}}^{(x)}_{n,l}$ to minimize the distance between two distributions, used for model training, i.e.,
\begin{align}
    \mathcal L=\frac{1}{NL} \sum\nolimits_{n=1}^{N} \sum\nolimits_{l=1}^{L} CrossEntropy(\bm{\tilde{p}}^{(x)}_{n,l},\bm{p}^{(x)}_{n,l}) \nonumber \\
    =-\frac{1}{NL} \sum\nolimits_{n=1}^{N} \sum\nolimits_{l=1}^{L} \log{\bm{\tilde{p}}^{(x)}_{n,l}(d^{(x)}_{n,l})},
\end{align}
where $\bm{\tilde{p}}^{(x)}_{n,l}(d^{(x)}_{n,l})$ denotes the $d^{(x)}_{n,l}$-th element in $\bm{\tilde{p}}^{(x)}_{n,l}$.

\section{Experiments}
\label{sec:exp}
\subsection{Dataset Construction and Task Definition}

We constructed clean speech dataset from the VoiceBank corpus \cite{valentini2016investigating}, with 23,075 utterances from 56 speakers for training set and 824 utterances from 2 unseen speakers for test set. 
The speech sampling rate is 44.1 kHz. 
We constructed task-specific degraded datasets based on a clean speech dataset, incorporating three conventional distortions (i.e., noise, reverberation, and band-limiting), three unconventional distortions (i.e., clipping, phase distortion, and compression distortion), and three mixed distortions. 
Each task is defined as follows.

1) \textbf{Denoising (DN)}: 
This task aims to enhance the degraded speech caused by additive noise. 
We constructed the noisy dataset by adding noise from the DEMAND dataset \cite{thiemann2013diverse} to the clean speech. 
For the training set, we used 5 types of noise, with the signal-to-noise ratio (SNR) ranging from 0 to 15 dB, in 5 dB intervals. 
For the test set, we used 5 types of unseen noise, with the SNR ranging from 2.5 to 17.5 dB in 5 dB intervals.

2) \textbf{Dereverberation (DR):} 
This task aims to enhance the degraded speech caused by reverberation. 
We adopted the room impulse response (RIR) dataset from the DNS Challenge \cite{dubey2024icassp}, which included 248 real RIRs and approximately 60,000 simulated RIRs. 
We constructed the reverberant dataset by randomly selecting RIRs and convolving RIRs with clean speech. 
The RIRs of the test set were unseen in the training set.

3) \textbf{Bandwidth Extension (BWE):}
This task aims to enhance the degraded speech caused by band-limiting. 
We constructed the band-limited dataset by downsampling the clean wideband speech to a 2 kHz sampling rate for both training and testing.

4) \textbf{Declipping (DC):}
This task aims to enhance the degraded speech caused by clipping.
We constructed the clipping dataset by restricting the amplitude of the original clean speech waveform to the range between 0.1 and 0.9 of their maximum values.

5) \textbf{Phase Distortion Restoration (PDR):} 
This task aims to enhance the degraded speech caused by inaccurate phase. 
We constructed the phase distortion dataset by replacing phase spectra of clean speech with the ones predicted by an NSPP\footnote{\url{https://github.com/YangAi520/NSPP}.} model \cite{ai2023neural}.

6) \textbf{Compression Distortion Restoration (CDR):} 
This task aims to enhance the degraded speech caused by compression from a speech codec. 
We constructed the compression distortion dataset by compressing clean speech using an APCodec\footnote{\url{https://github.com/YangAi520/APCodec}.} \cite{ai2024apcodec} with a single VQ.

\textbf{7) Mixed Distortion Restoration:} 
We combined the six types of distortions mentioned above to create three tasks, represented as \textbf{DN}+\textbf{DR}+\textbf{BWE}, which stands for denoising + dereverberation + bandwidth extension (band-limited speech sampling rate is 8 kHz), \textbf{DN}+\textbf{DR}+\textbf{DC}, which stands for denoising + dereverberation + declipping, and \textbf{DN}+\textbf{PDR}+\textbf{CDR}, which stands for denoising + phase distortion restoration + compression distortion restoration. 
We did not downsample the speech to 2 kHz in the \textbf{DN+DR+BWE} task, as in the standalone \textbf{BWE} task, because further adding noise and reverberation to 2 kHz speech would render the speech nearly unintelligible and the SE solution unrealistically ill-posed.

\begin{table*}[!htbp]
\centering
\caption{Objective evaluation results among UDSE and baselines for different SE tasks.}
\label{tab:results}
\resizebox{0.74\linewidth}{!}{
\begin{tabular}{cccccccccccc}
\hline
\multicolumn{6}{c}{SE Task} & \multirow{3}*{Model} & \multicolumn{5}{c}{Objective Metrics}\\
    \multirow{2}*{\textbf{DN}}& \multirow{2}*{\textbf{DR}}& \multirow{2}*{\textbf{BWE}}& \multirow{2}*{\textbf{DC}}& \multirow{2}*{\textbf{PDR}}& \multirow{2}*{\textbf{CDR}} &  & \multirow{2}*{\textbf{NISQA$\uparrow$}} & \multicolumn{3}{c}{\textbf{DNSMOS$\uparrow$}} & \multirow{2}*{\textbf{UTMOS$\uparrow$}} \\ 
    & & & & & & & & SIG & BAK & OVRL & \\
    \hline
     & & & & & & Clean Speech & 4.62 & 3.51& 4.04& 3.22  &  4.10 \\ \hline
    \multirow{6}*{\Checkmark}& \multirow{6}*{\XSolidBrush}& \multirow{6}*{\XSolidBrush}& \multirow{6}*{\XSolidBrush}& \multirow{6}*{\XSolidBrush}& \multirow{6}*{\XSolidBrush} & $\textbf{DEMUCS}$ & 3.45 & 3.37& 3.96& 3.07 & 3.63 \\  
     & & & & & & $\textbf{CMGAN}$  & 4.51 & \textbf{3.51}& 4.03& \textbf{3.21} & 3.99 \\ 
     & & & & & & $\textbf{MP-SENet}$ & 4.56 & \textbf{3.51}& \textbf{4.04}& \textbf{3.21} & \textbf{4.01} \\  
     & & & & & & $\textbf{UniverSE++}$ & 4.44 & 3.46 & 4.03 & 3.17 & 3.88 \\
     & & & & & & $\textbf{Genhancer}$ & 3.97 & 3.27 & 3.94 & 2.96 & 3.43 \\
     & & & & & & $\textbf{UDSE}$ & \textbf{4.60} & 3.49& 3.99& 3.18 & 3.88 \\ 
     \hline
    \multirow{6}*{\XSolidBrush}& \multirow{6}*{\Checkmark}& \multirow{6}*{\XSolidBrush}& \multirow{6}*{\XSolidBrush}& \multirow{6}*{\XSolidBrush} &\multirow{6}*{\XSolidBrush} & \textbf{DEMUCS} & 1.43 & 2.69& 3.91& 2.44 & 1.30 \\ 
     & & & & & & \textbf{CMGAN} & 1.13 & 2.62& 3.99& 2.28 & 1.55 \\ 
     & & & & & & \textbf{MP-SENet} & 4.34 & \textbf{3.50}& \textbf{4.06}& \textbf{3.21} & 3.65 \\ 
     & & & & & & $\textbf{UniverSE++}$ & 3.90 & 3.02 & 3.97 & 2.74 & 2.59 \\
     & & & & & & $\textbf{Genhancer}$ & 2.05 & 2.01 & 3.46 & 1.83 & 1.68 \\
     & & & & & & \textbf{UDSE} & \textbf{4.60} & \textbf{3.50}& 4.01& 3.19 & \textbf{3.82} \\ 
    \hline
    \multirow{7}*{\XSolidBrush}& \multirow{7}*{\XSolidBrush}& \multirow{7}*{\Checkmark}& \multirow{7}*{\XSolidBrush}& \multirow{7}*{\XSolidBrush} &\multirow{7}*{\XSolidBrush} & \textbf{DEMUCS}  & 1.91 & 3.30& 3.97& 2.99 & 1.80 \\ 
     & & & & & & \textbf{CMGAN} & 2.57 & 3.35& 3.98& 3.03 & 2.94 \\ 
     & & & & & & \textbf{MP-SENet} & 1.59 & 3.26& 3.95& 2.94 & 2.38 \\ 
     & & & & & & \textbf{AP-BWE} & 4.04 & 3.45& 4.01& 3.15 & 3.17\\ 
     & & & & & & $\textbf{UniverSE++}$ & 3.79 & 3.39 & 3.99 & 3.08 & 3.50 \\
     & & & & & & $\textbf{Genhancer}$ & 3.44 & 3.25 & 3.93 & 2.94 & 3.01 \\
     & & & & & & \textbf{UDSE} & \textbf{4.48} & \textbf{3.49}& \textbf{4.03}& \textbf{3.19} & \textbf{3.87} \\ 
    \hline
    \multirow{6}*{\XSolidBrush}& \multirow{6}*{\XSolidBrush}& \multirow{6}*{\XSolidBrush}& \multirow{6}*{\Checkmark}& \multirow{6}*{\XSolidBrush} &\multirow{6}*{\XSolidBrush} & \textbf{DEMUCS} & 3.11 & 3.49 & 4.03 & 3.19 & 3.79 \\ 
     & & & & & & \textbf{CMGAN} & \textbf{4.57} & 3.49 & 4.03 & 3.20 & 4.04 \\ 
     & & & & & & \textbf{MP-SENet} & 4.53 & \textbf{3.51} & 4.04 & \textbf{3.22} & \textbf{4.08} \\ 
     & & & & & & $\textbf{UniverSE++}$ & 3.98 & 3.48 & \textbf{4.05} & 3.19 & 3.94 \\
     & & & & & & $\textbf{Genhancer}$ & 3.81 & 3.26 & 3.96 & 2.96 & 3.14 \\
     & & & & & & \textbf{UDSE} & 4.49 & 3.50 & 4.03 & 3.21 & 4.03 \\ 
    \hline
    \multirow{6}*{\XSolidBrush}& \multirow{6}*{\XSolidBrush}& \multirow{6}*{\XSolidBrush}& \multirow{6}*{\XSolidBrush}& \multirow{6}*{\Checkmark}& \multirow{6}*{\XSolidBrush} & \textbf{DEMUCS} & 2.03 & 3.04& 3.98& 2.77 & 2.55 \\ 
     & & & & & & \textbf{CMGAN} & 4.43 & 3.32& 3.98& 3.02 & 3.74 \\ 
     & & & & & & \textbf{MP-SENet} & \textbf{4.64} & 3.48& \textbf{4.04}& 3.19 & \textbf{4.04} \\ 
     & & & & & & $\textbf{UniverSE++}$ & 4.59 & 3.44 & 4.03 & 3.15 & 3.95 \\
     & & & & & & $\textbf{Genhancer}$ & 4.15 & 3.27 & 3.95 & 2.97 & 3.44 \\
     & & & & & & \textbf{UDSE} & \textbf{4.64} & \textbf{3.52}& 4.03& \textbf{3.22} & 4.02 \\ 
    \hline
    \multirow{6}*{\XSolidBrush}& \multirow{6}*{\XSolidBrush}& \multirow{6}*{\XSolidBrush}& \multirow{6}*{\XSolidBrush}& \multirow{6}*{\XSolidBrush}& \multirow{6}*{\Checkmark} & \textbf{DEMUCS} & 1.99 & 3.14& 3.99& 2.85 & 2.28 \\ 
     & & & & & & \textbf{CMGAN} & 3.26 & 3.28& 3.99& 2.97  & 3.25 \\
     & & & & & & \textbf{MP-SENet} & 2.47 & 3.36& 4.02& 3.06  & 3.14 \\ 
     & & & & & & $\textbf{UniverSE++}$ & 4.35 & 3.35 & \textbf{4.03} & 3.06 & 3.60 \\
     & & & & & & $\textbf{Genhancer}$ & 4.31 & 3.22 & 3.96 & 2.93 & 3.19 \\
     & & & & & & \textbf{UDSE} & \textbf{4.67} & \textbf{3.49}& \textbf{4.03}& \textbf{3.20} & \textbf{3.90} \\ 
    \hline
    \multirow{6}*{\Checkmark}& \multirow{6}*{\Checkmark}& \multirow{6}*{\Checkmark}& \multirow{6}*{\XSolidBrush}& \multirow{6}*{\XSolidBrush}& \multirow{6}*{\XSolidBrush} & \textbf{DEMUCS} & 0.77 & 1.83& 3.54& 1.51 & 1.33 \\ 
     & & & & & & \textbf{CMGAN} & 0.85 & 2.30& 3.71& 1.94 & 1.38 \\ 
     & & & & & & \textbf{MP-SENet} & 2.33 & 2.79& 3.80& 2.46 & 2.20 \\
     & & & & & & $\textbf{UniverSE++}$ & 3.89 & 2.95 & 3.97 & 2.68 & 2.51 \\
     & & & & & & $\textbf{Genhancer}$ & 2.05 & 2.13 & 3.62 & 1.93 & 1.65 \\
     & & & & & & \textbf{UDSE} & \textbf{4.37} & \textbf{3.46}& \textbf{4.01}& \textbf{3.16} & \textbf{3.60} \\ 
    \hline
    \multirow{6}*{\Checkmark}& \multirow{6}*{\Checkmark}& \multirow{6}*{\XSolidBrush}& \multirow{6}*{\Checkmark}& \multirow{6}*{\XSolidBrush}& \multirow{6}*{\XSolidBrush} & \textbf{DEMUCS} & 0.78 & 1.86 & 3.51 & 1.51 & 1.34 \\ 
     & & & & & & \textbf{CMGAN} & 1.04 & 2.05 & 3.39 & 1.73 & 1.34 \\ 
     & & & & & & \textbf{MP-SENet} & 1.54 & 3.11 & 3.94 & 2.79 & 1.93 \\ 
     & & & & & & $\textbf{UniverSE++}$ & 3.74 & 2.92 & 3.97 & 2.65 & 2.36 \\
     & & & & & & $\textbf{Genhancer}$ & 2.00 & 2.11 & 3.62 & 1.92 & 1.62 \\
     & & & & & & \textbf{UDSE} & \textbf{4.25} & \textbf{3.43} & \textbf{4.01} & \textbf{3.13} & \textbf{3.47} \\ 
    \hline
    \multirow{6}*{\Checkmark}& \multirow{6}*{\XSolidBrush}& \multirow{6}*{\XSolidBrush}& \multirow{6}*{\XSolidBrush}& \multirow{6}*{\Checkmark}& \multirow{6}*{\Checkmark} & \textbf{DEMUCS} & 1.92 & 2.92& 4.05& 2.68 & 1.65 \\  
     & & & & & & \textbf{CMGAN} & 2.93 & 3.29& 3.85& 2.93 & 2.96 \\ 
     & & & & & & \textbf{MP-SENet} & 2.26 & 3.38& 4.05& 3.08  & 2.62  \\ 
     & & & & & & $\textbf{UniverSE++}$ & 4.41 & 3.33 & 4.06 & 3.05 & 3.52 \\
     & & & & & & $\textbf{Genhancer}$ & 3.87 & 3.13 & 4.04 & 2.89 & 2.87 \\
     & & & & & & \textbf{UDSE} & \textbf{4.57}  & \textbf{3.56}& \textbf{4.09}& \textbf{3.29} & \textbf{3.96} \\ \hline      
\end{tabular}}
\end{table*}

\subsection{Model Details}

The UDSE\footnote{Codes and speech samples are available at: \url{https://fliu215.github.io/UDSE/}.} utilized a DAC\footnote{\url{https://github.com/descriptinc/descript-audio-codec}.} \cite{kumar2023high} as $\phi_{Codec}$, which included 9 VQs (i.e., $N=9$), with both the codebook size and the code vector dimension set to 1024 (i.e., $K=M=1024$). 
The feature processor $\phi_{FP}$ used 8 Conformer blocks (i.e., $B_G=8$), while each token predictor $\phi_{TP_1},\dots,\phi_{TP_N}$ used 4 Conformer blocks (i.e., $B_T=4$). 
The number of channels and attention heads in each block were 512 and 8, respectively (i.e., $C=512$). 
We trained the UDSE using an AdamW optimizer on a single {NVIDIA} A800 GPU, with $\beta_{1}=0.9,\beta_{2}=0.95,$ and a weight decay of 0.01 for 100 epochs. 
The initial learning rate was set to 0.0005, with a cosine annealing strategy for decay and a warm-up training scheduler for the first 4k steps.

We compared UDSE with several advanced continuous-domain SE {models}, including waveform-regression-based DEMUCS\footnote{\url{https://github.com/facebookresearch/denoiser}.} \cite{defossez2020real}, spectrum-regression-based CMGAN\footnote{\url{https://github.com/ruizhecao96/CMGAN}.} \cite{abdulatif2024cmgan} and MP-SENet\footnote{\url{https://github.com/yxlu-0102/MP-SENet}.} \cite{lu2023mp,lu2025explicit}, and diffusion-based UniverSE++\footnote{\url{https://github.com/line/open-universe}.} \cite{universe++}. 
In addition, we have also reproduced the Genhancer \cite{yang2024genhancer} as a discrete-domain baseline. 
For task \textbf{BWE}, we also compared UDSE with the advanced continuous-domain AP-BWE\footnote{\url{https://github.com/yxlu-0102/AP-BWE}.} \cite{lu2024towards}, specifically designed for speech bandwidth extension.
Except for Genhancer, we reproduced all the models using the official codes they provided, whereas for Genhancer, we had to reproduce it manually due to the lack of open-source code.
All models were conducted on our 44.1 kHz dataset.

\begin{table*}[t]
\centering
\caption{Average preference scores (\%) of ABX subjective tests on speech quality between UDSE and a baseline (i.e., model B) on different SE tasks, where N/P stands for ``no preference" and $p$ denotes the $p$-value of a $t$-test between two models. For tasks \textbf{DN}, \textbf{DR}, \textbf{DC} and \textbf{PDR}, ``model B" is MP-SENet, and {for tasks \textbf{CDR}, \textbf{DN}+\textbf{DR}+\textbf{BWE}, \textbf{DN}+\textbf{DR}+\textbf{DC} and \textbf{DN}+\textbf{PDR}+\textbf{CDR}, ``model B" is UniverSE++}, while for task \textbf{BWE}, ``model B" is AP-BWE.}
\label{tab:ABX}
\begin{tabular}{cccccccccc}
\hline
\multicolumn{6}{c}{SE Task} & \multirow{2}*{\textbf{UDSE}} & \multirow{2}*{Model B} & \multirow{2}*{N/P} & \multirow{2}*{$p$}\\
    \textbf{DN}& \textbf{DR}& \textbf{BWE}& \textbf{DC}& \textbf{PDR}& \textbf{CDR} &  &  &  &  \\ 
    \hline
    \Checkmark& \XSolidBrush& \XSolidBrush& \XSolidBrush& \XSolidBrush& \XSolidBrush & 41.96 & 41.25 & 16.79 & 0.85 \\  
    \XSolidBrush& \Checkmark& \XSolidBrush& \XSolidBrush& \XSolidBrush& \XSolidBrush & \textbf{56.92} & 36.35 & 6.73 & \textbf{$\bm{<}$ 0.01} \\  
    \XSolidBrush& \XSolidBrush& \Checkmark& \XSolidBrush& \XSolidBrush& \XSolidBrush & \textbf{54.60} & 38.60 & 6.80 & \textbf{$\bm{<}$ 0.01} \\ 
    \XSolidBrush& \XSolidBrush& \XSolidBrush& \Checkmark& \XSolidBrush& \XSolidBrush & {26.35} & {32.50} & 41.15 & {0.07} \\
    \XSolidBrush& \XSolidBrush& \XSolidBrush& \XSolidBrush& \Checkmark& \XSolidBrush & \textbf{46.83} & 37.01 & 16.16 & \textbf{$\bm{<}$ 0.01} \\ 
    \XSolidBrush& \XSolidBrush& \XSolidBrush& \XSolidBrush& \XSolidBrush& \Checkmark & \textbf{79.93} & 12.65 & 7.42 & \textbf{$\bm{<}$ 0.01} \\ 
    \Checkmark& \Checkmark& \Checkmark& \XSolidBrush& \XSolidBrush& \XSolidBrush & \textbf{83.52} & 6.11 & 10.37 & \textbf{$\bm{<}$ 0.01} \\ 
    \Checkmark& \Checkmark& \XSolidBrush& \Checkmark& \XSolidBrush& \XSolidBrush & \textbf{86.55} & 8.28 & 5.17 & \textbf{$\bm{<}$ 0.01} \\
    \Checkmark& \XSolidBrush& \XSolidBrush& \XSolidBrush & \Checkmark& \Checkmark & \textbf{66.67} & 24.50 & 8.83 & \textbf{$\bm{<}$ 0.01} \\ 
 \hline      
\end{tabular}
\end{table*}

\begin{figure*}[t]
    \centering
    \includegraphics[width=\linewidth]{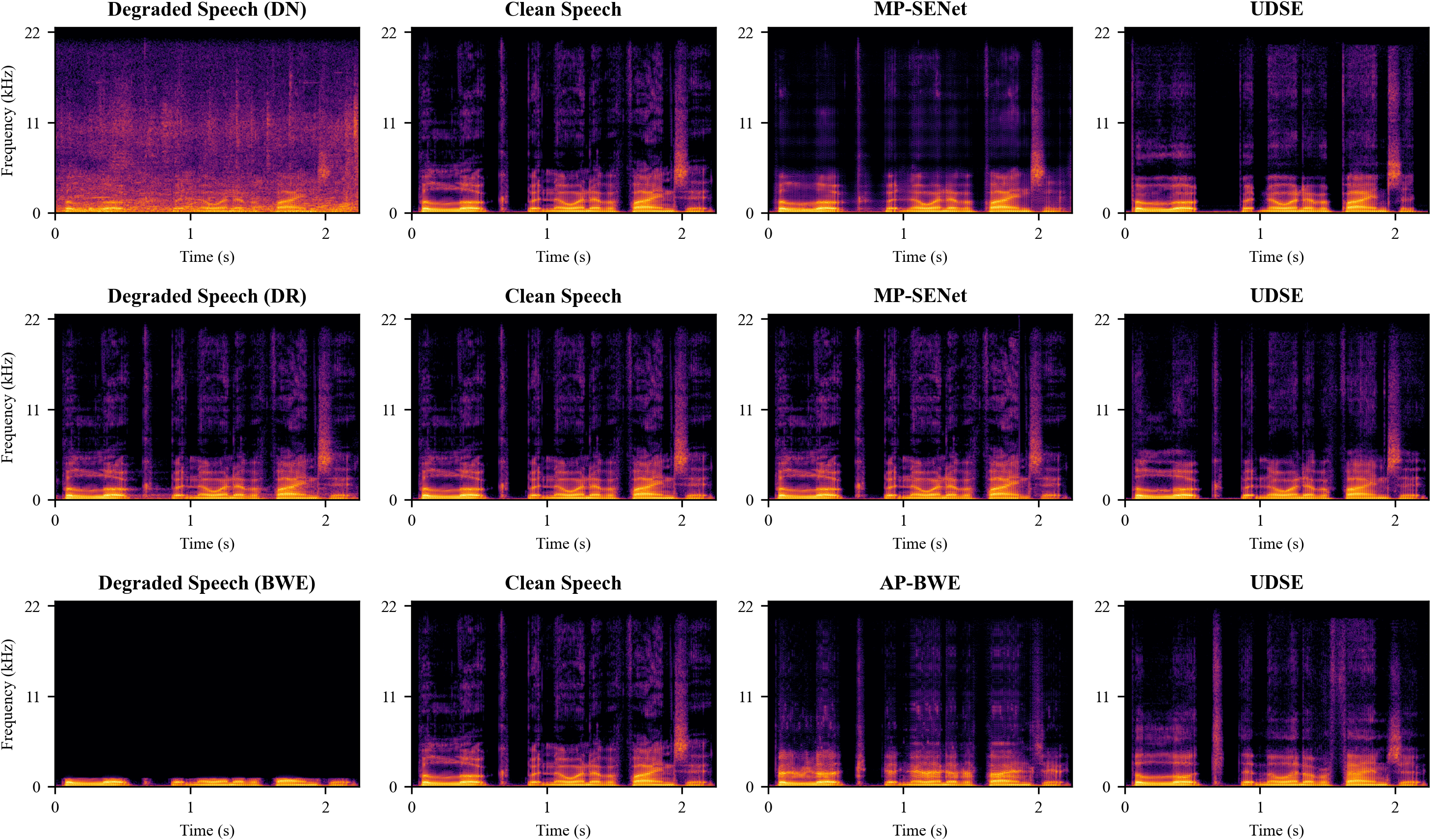}
    \caption{Spectrogram comparison among degraded speech, clean speech, and speeches enhanced by the baseline with the best objective scores and UDSE for conventional \textbf{DN}, \textbf{DR} and \textbf{BWE} tasks, respectively.}
    \label{fig:spectrum1}
\end{figure*}

\subsection{Evaluation Metrics}
\label{Evaluation Metrics}

For objective evaluation, since UDSE predicts the distribution of discrete tokens corresponding to clean speech (i.e., it formulates SE as a classification task rather than directly generating the clean speech as in regression-based models), traditional intrusive indicators such as perceptual evaluation of speech quality (PESQ) are not well-suited for evaluating our model, as suggested in \cite{yao2025gense}.
Therefore, we used three non-intrusive metrics commonly adopted in SE field, including NISQA \cite{mittag2021nisqa}, DNSMOS \cite{reddy2022dnsmos} and UTMOS \cite{saeki2022utmos}, to assess overall speech quality. 
NISQA is a full-band metric that can evaluate the overall quality of 44.1 kHz speech, while DNSMOS and UTMOS are both designed for 16 kHz, focus more on the low-frequency quality. 
All three metrics use the same scoring scale as traditional mean opinion score (MOS), ranging from 1 to 5. 
Both NISQA and UTMOS directly reflect the perceived acceptability and naturalness of speech. DNSMOS, on the other hand, follows the ITU-T P.835 subjective test framework and provides multi-dimensional scores: SIG represents speech quality and primarily reflects overall signal distortion; BAK measures the perceptual intensity of background noise, with higher scores indicating better noise suppression; and OVRL provides an overall rating, indicating the general naturalness of the speech.

For the subjective evaluation, we conducted ABX preference tests on the Amazon Mechanical Turk\footnote{\url{https://www.mturk.com/}.} platform to compare the differences between UDSE and the baseline model with the best objective performance. 
In each ABX test, 20 speech samples enhanced by both compared models were randomly selected from the test set.
These samples were evaluated by at least 30 native English-speaking listeners. 
The listeners were asked to judge which of the two speech samples in each pair had better speech quality or whether there was no preference. 
In addition to calculating the average preference score, we also used the $p$-value from a $t$-test to assess the statistical significance of the differences between the two models.

\begin{figure*}[t]
    \centering
    \includegraphics[width=\linewidth]{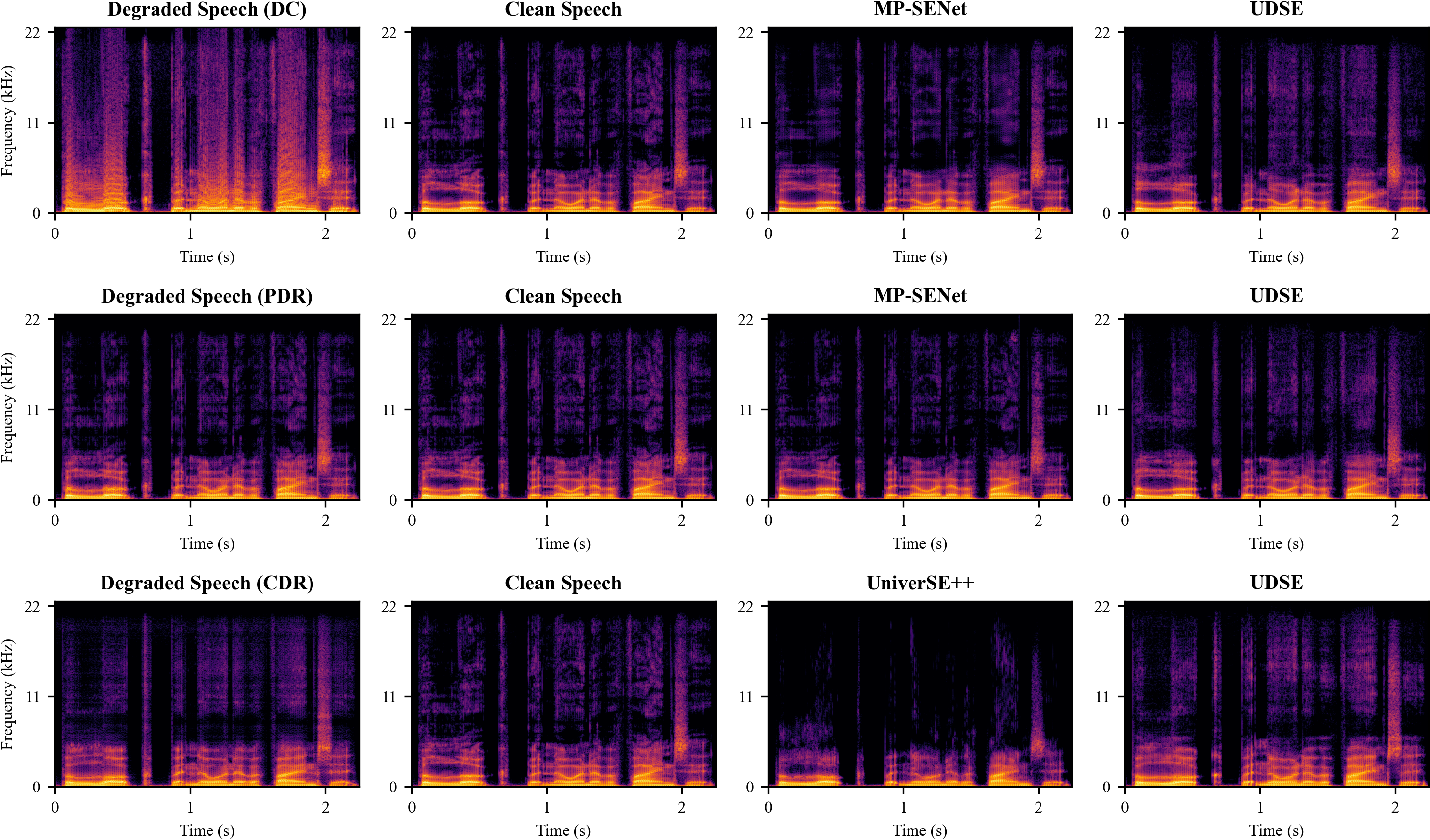}
    \caption{{Spectrogram comparison among degraded speech, clean speech, and speeches enhanced by the baseline with the best objective scores and UDSE for unconventional \textbf{DC}, \textbf{PDR} and \textbf{CDR} tasks, respectively.}}
    \label{fig:spectrum2}
\end{figure*}

\subsection{Preliminary Experimental Results}

\begin{figure*}[t]
    \centering
    \includegraphics[width=\linewidth]{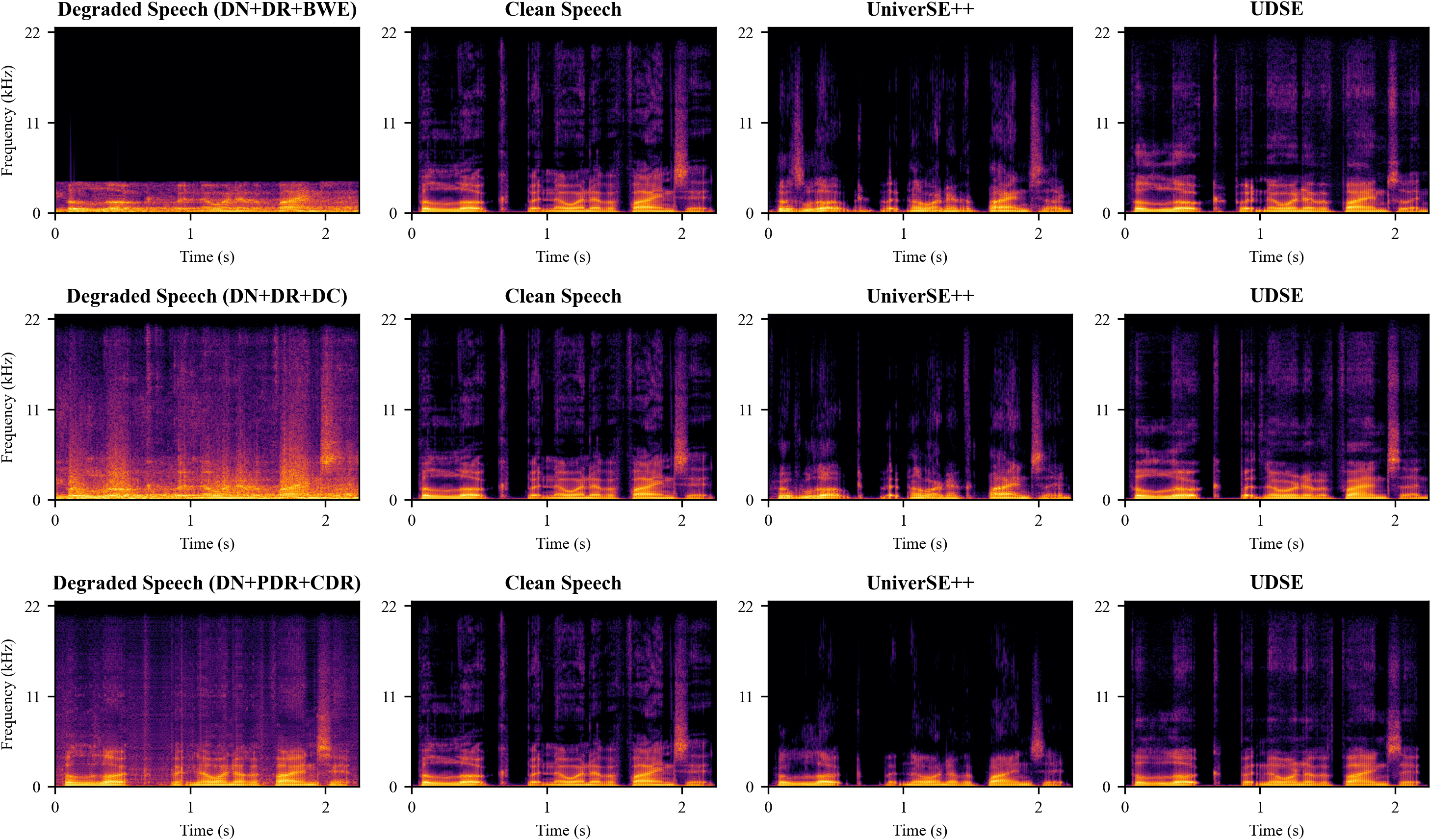}
    \caption{{Spectrogram comparison among degraded speech, clean speech, and speeches enhanced by the baseline with the best objective scores and UDSE for mixed \textbf{DN}+\textbf{DR}+\textbf{BWE}, \textbf{DN}+\textbf{DR}+\textbf{DC} and \textbf{DN}+\textbf{PDR}+\textbf{CDR} tasks, respectively.}}
    \label{fig:spectrum3}
\end{figure*}

The objective and subjective evaluation results are shown in Tables \ref{tab:results} and \ref{tab:ABX}, respectively. 
To provide an intuitive visualization of the enhancement performance, we also drew the spectrograms of degraded speech, clean speech, and speech enhanced by both the baseline model with the best objective scores and UDSE across three task types (i.e., conventional, unconventional and mixed), as shown in Figures \ref{fig:spectrum1}, \ref{fig:spectrum2}, and \ref{fig:spectrum3}, respectively. 

\subsubsection{Experimental Results for Conventional Distortions}

For the classic speech denoising task \textbf{DN}, our proposed UDSE achieved objective results comparable to the advanced continuous-domain baseline models, as shown in Table \ref{tab:results}. 
It can be seen that UDSE outperformed all baselines in NISQA scores and was similar to MP-SENet and CMGAN in the three indicators of DNSMOS. 
However, the UTMOS score of UDSE was slightly lower than that of MP-SENet and CMGAN. 
This may be attributed to the fact that UTMOS can only evaluate the performance of speech in the low-frequency band (0$\sim$8 kHz). 
In conjunction with the fullband NISQA results, it can be inferred that the proposed UDSE possesses a stronger ability to reconstruct the high-frequency components of speech. 
This is also verified in the first row of spectrograms in Figure \ref{fig:spectrum1}, where it is clearly observed that MP-SENet struggles to recover high-frequency information (8 kHz$\sim$22.05 kHz) under low-SNR noisy conditions, while UDSE successfully preserves these components. 
Considering subjective perception, the ABX preference test results presented in Table \ref{tab:ABX} indicate no significant perceptual difference between the noisy speech enhanced by UDSE and that enhanced by MP-SENet ($p=0.85$). 
These findings suggest that the proposed discrete-domain UDSE model can perform on par with continuous-domain models for \textbf{DN} task.


For the classic dereverberation task \textbf{DR}, our proposed UDSE significantly outperformed the advanced continuous-domain baseline models in most objective metrics, as shown in Table \ref{tab:results}. 
Specifically, UDSE significantly outperformed all baselines in both NISQA and UTMOS scores, and achieved a comparable DNSMOS score to MP-SENet, effectively suppressing speech distortion caused by reverberation. 
It was evident that although CMGAN performed well in denoising tasks, it lagged behind in all objective metrics when dealing with dereverberation. 
In contrast, the proposed UDSE consistently achieved robust performance across both distortion types. 
The second row of Figure \ref{fig:spectrum1} shows the spectrograms of reverberant speech enhanced by MP-SENet and UDSE under low reverberation time conditions. 
Interestingly, the spectrogram of the speech enhanced by MP-SENet is more similar to that of the clean reference speech. 
This may be attributed to the fact that regression-based continuous-domain SE models directly model the speech waveform or features. 
In particular, MP-SENet explicitly predicts the amplitude spectra, which makes the enhanced speech more ``similar” to the reference speech, especially when the distortion is not severe (e.g., under high SNR or low reverberation time conditions). 
In contrast, classification-based discrete-domain models indirectly predict discrete tokens of speech and then decode them into speech waveforms, which may result in enhanced speech that is ``not similar” to the reference speech. 
This is also why intrusive indicators that rely on reference speech are not suitable for evaluating discrete-domain models, as stated in Section \ref{Evaluation Metrics}. 
At this point, subjective evaluation becomes the gold standard for assessing speech quality. 
As shown in the results in Table \ref{tab:ABX}, this lack of ``similarity” does not negatively impact perceptual quality — reverberant speech enhanced by UDSE is even significantly better than that enhanced by MP-SENet ($p<0.01$).


{For the bandwidth extension task \textbf{BWE}, as suggested in Table \ref{tab:results}, DEMUCS, CMGAN, MP-SENet and Genhancer, which were specifically designed for denoising and reverberation, all failed. 
The more general-purpose UniverSE++ also performed poorly.}
Although the objective results of AP-BWE, which was specifically designed for task \textbf{BWE}, outperformed the five models mentioned above, there is still a significant gap compared to our proposed UDSE. 
Subjectively, UDSE also significantly outperformed AP-BWE ($p<0.01$) as shown in Table \ref{tab:ABX}. 
According to listener feedback, the speech extended by AP-BWE contained noticeable pronunciation errors, whereas UDSE did not. 
The last row of Figure \ref{fig:spectrum1} intuitively shows the results of bandwidth extension. 
It is evident that UDSE achieves a richer and more accurate reconstruction of high-frequency speech details compared to AP-BWE. 
For example, within the time range of 1 to 1.2 seconds and the frequency band of 1 kHz to 6 kHz, the harmonic details of the speech extended by UDSE are more similar to those of the clean speech. 
This indicates that UDSE has potential in speech bandwidth extension for extremely narrow bands (e.g., from 2 kHz to 44.1 kHz, representing an expansion of over 22 times in bandwidth).

\subsubsection{Experimental Results for Unconventional Distortions}

For the declipping task \textbf{DC}, the results in Table \ref{tab:results} showed that for this type of distortion, both our proposed UDSE and the baseline models achieved satisfactory performance, with only minor differences in the scores across various objective metrics. 
As demonstrated in the first row of Figure \ref{fig:spectrum2}, the observation is consistent with the \textbf{DR} task: UDSE produces declipped speech that is slightly less similar to the reference than MP-SENet. 
Nevertheless, subjective evaluation results presented in Table \ref{tab:ABX} reveal no significant perceptual difference between the two approaches ($p=0.07$). 
This indicates that clipping is a relatively simple type of distortion, which can be effectively addressed by most SE approaches.


{For the phase distortion restoration task \textbf{PDR}, UDSE achieved objective results comparable to those of MP-SENet, while outperforming DEMUCS, CMGAN, UniverSE++ and Genhancer, as shown in Table \ref{tab:results}.
MP-SENet, by introducing explicit phase prediction, was more suitable for phase distortion recovery compared to DEMUCS, CMGAN, UniverSE++ and Genhancer.}  
As shown in the second row of Figure \ref{fig:spectrum2}, speech affected by phase distortion exhibited prominent horizontal stripes — perceptually experienced as an irritating buzzing sound.
In contrast, the spectrogram of the UDSE-enhanced speech showed no such patterns, indicating that UDSE effectively eliminated interference caused by phase distortion.
Since phase restoration effects are difficult to visualize clearly in spectrograms, subjective evaluation is necessary to provide additional evidence.
As shown in the subjective evaluation results in Table \ref{tab:ABX}, UDSE-enhanced speech was generally preferred by listeners ($p<0.01$).

For the compression distortion restoration task \textbf{CDR}, our proposed UDSE significantly outperformed all baseline models in both objective and subjective aspects ($p<0.01$), as shown in Tables \ref{tab:results} and \ref{tab:ABX}. 
Interestingly, we found that these baseline models even had a negative effect on the enhancement of compressed speech. 
As clearly shown in the last row of Figure \ref{fig:spectrum2}, UDSE effectively restored spectral details, with the enhanced harmonics closely resembling those of the clean speech.
{Although the speech enhanced by UniverSE++ showed decent restoration of the low-frequency harmonics, it lost many spectral details in the high-frequency regions, which significantly degraded the listening experience. 
These results suggest that baseline SE models are less effective in handling codec-induced compression distortions, likely due to the strong correlation between compressed and clean speech signals.} 
In comparison, the proposed UDSE tackles this challenge from a discrete-domain perspective.

\subsubsection{Experimental Results for Mixed Distortions}

{In the three mixed distortion restoration tasks, our proposed UDSE had a significant advantage over all baseline models in both objective and subjective aspects ($p<0.01$), as shown in Tables \ref{tab:results} and \ref{tab:ABX}. 
For task \textbf{DN}+\textbf{DR}+\textbf{BWE}, UDSE outperformed UniverSE++ by more than 1-point in the UTMOS score. 
In addition, for task \textbf{DN}+\textbf{DR}+\textbf{DC}, UDSE achieved a 0.4-point higher OVRL score in DNSMOS and a 1.1-point higher UTMOS score compared to UniverSE++. 
Consistently, for task \textbf{DN}+\textbf{PDR}+\textbf{CDR}, UDSE again exceeded UniverSE++ by over 0.4-point in the UTMOS metric. 
Subjective evaluations showed that, in all three mixed-distortion tasks, the number of participants preferring UDSE-enhanced speech was at least double that of those favoring UniverSE++ — highlighting UDSE’s superior performance in managing complex distortions.}

\begin{table*}[t]
\centering
\caption{Average preference scores (\%) of ABX subjective tests on speech quality between UDSE and its variants on \textbf{DN}+\textbf{DR}+\textbf{BWE} task, where N/P stands for ``no preference" and $p$ denotes the $p$-value of a $t$-test between two models. }
\label{tab:a}
\begin{tabular}{ccccccc}
\hline
 {\textbf{UDSE}} & \textbf{UDSE w/ Transformer} & \textbf{UDSE w/ Parallel Mode} & \textbf{UDSE w/o Global Condition} & \textbf{UDSE w/ MDCTCodec} & {N/P} & {$p$}\\ \hline
 \textbf{73.00} & 25.33 & - & - & - & 1.67 & \textbf{$\bm{<}$ 0.01} \\  
 \textbf{74.83} & - & 24.31 & - & - & 0.86 & \textbf{$\bm{<}$ 0.01} \\  
 \textbf{58.00} & - & - & 40.00 & - & 2.00 & \textbf{$\bm{<}$ 0.01} \\ 
  47.87 & - & - & - & 49.50 & 2.63 & 0.64 \\ 
 \hline      
\end{tabular}
\end{table*}

The spectrograms shown in Figure \ref{fig:spectrum3} further provide a clear visual illustration of the effectiveness of the proposed UDSE in enhancing speech with mixed distortions.  
{We can observe that for the \textbf{DN}+\textbf{DR}+\textbf{BWE} task (the first row in Figure \ref{fig:spectrum3}), UDSE recovered the high-frequency components much more accurately, bringing the result closer to the ground-truth clean speech, while UniverSE++ struggled with recovering the high-frequency part. 
For the \textbf{DN}+\textbf{DR}+\textbf{DC} task (the second row in Figure \ref{fig:spectrum3}), the speech enhanced by UniverSE++ lost a lot of detail in the high-frequency range and had a noticeable fundamental frequency error around 1 second, resulting in a poor listening experience.
For the \textbf{DN}+\textbf{PDR}+\textbf{CDR} task (the last row in Figure \ref{fig:spectrum3}), the spectrogram of the speech recovered by UDSE retained more details, while UniverSE++ produced significant high-frequency loss and poor spectral detail recovery, significantly impacting the listening experience. }
All the above experimental results demonstrate that the proposed UDSE can effectively mitigate various types of distortion in degraded speech, making it more suitable for real-world applications.

In summary, our proposed UDSE can not only handle various single distortions but also effectively address mixed distortions.
This confirms the universality of UDSE and the advantages of using discrete-domain classification solutions for SE, compared to regression-based continuous-domain models.

\subsection{Analysis and Discussion}
To further investigate the contribution of each component in the proposed UDSE model, we conducted a series of analytical experiments. 
For simplicity, the experiments were conducted under a commonly encountered mixed distortion condition, i.e., \textbf{DN}+\textbf{DR}+\textbf{BWE} task. 
To maintain a concise evaluation process, only subjective preference listening tests were carried out to assess the perceptual differences between UDSE and its variants.
The results of the subjective evaluation are presented in Table \ref{tab:a}.

\subsubsection{Backbone Network Selection Analysis}
In the UDSE framework, we utilized the Conformer block as the core backbone network to support both feature processing and token predictions.
To investigate the impact of different backbone networks on UDSE's performance, we replaced the Conformer blocks with Transformer blocks (denoted as UDSE w/ Transformer), which also incorporated multi-head self-attention. 
The results in Table \ref{tab:a} clearly indicate that UDSE with a Transformer backbone performed significantly worse than the original UDSE ($p<0.01$).
This may be attributed to the fact that the Transformer only models global dependencies, whereas the Conformer enhances representational capacity by integrating convolutional layers for local feature extraction with self-attention mechanisms for capturing global context.
Therefore, the Conformer, which captures both local and global features, is likely more suitable for our token prediction task than the Transformer. 
This advantage in representational capacity is the primary reason we selected it as the backbone network in UDSE.

\subsubsection{Token Prediction Mode Analysis}
In the UDSE framework, we proposed a novel token prediction mode based on the RVQ quantization rule, where clean tokens quantized by each VQ are predicted sequentially, with each prediction conditioned on all preceding prediction results.
To evaluate the effectiveness of this token prediction mode, we conducted an ablation study in which the prediction of the current VQ's clean token was performed without relying on the outputs of previously predicted tokens. 
In this case, the clean tokens quantized by VQs were predicted in parallel by token predictors $\phi_{TP_1},\dots,\phi_{TP_N}$, with no interconnections among them (denoted as UDSE w/ Parallel Mode).
The results in Table \ref{tab:a} show that UDSE significantly outperformed UDSE w/ Parallel Mode ($p<0.01$). 
This performance gap indicates that the RVQ-based sequential prediction mode used in UDSE is beneficial, as it effectively leverages the correlations among different VQ stages. 
The parallel prediction mode, by ignoring these dependencies, leads to a notable degradation in enhancement performance.
These experimental findings validate the effectiveness of the proposed sequential token prediction mode based on the RVQ rule.

\subsubsection{Global Condition Analysis}
In the UDSE framework, to reduce the difficulty of token prediction, global features extracted by a dedicated module are used as global conditioning for the prediction of each VQ. 
Given the characteristics of RVQ, the information provided to the first token predictor is naturally propagated to subsequent predictors. 
This raises an important question: can the global features be effectively treated as local, such that they only need to be provided to the first token predictor? 
To verify this, we constructed a variant of UDSE (denoted as UDSE w/o Global Condition) in which the global features are provided only to the first token predictor $\phi_{TP_1}$ (i.e., removing $\bm{G}^{(y)}$ from the second line of Equation (\ref{a})).
The results in Table \ref{tab:a} show that removing the global conditioning leads to a significant decline ($p<0.01$) in the subjective quality of the enhanced speech produced by UDSE. 
This indicates that, although the first token predictor is conditioned on features extracted from the degraded speech, these features become increasingly diluted as they propagate through the network, ultimately reducing the effectiveness of subsequent token predictors. 
Providing global conditioning features explicitly to each token predictor is essential for preserving their prediction capabilities, and plays a critical role in ensuring the overall performance of UDSE.


\subsubsection{Codec Generalization Analysis}
In the UDSE framework, the neural speech codec is a crucial component, serving multiple roles including extracting global feature condition, providing clean token targets for training, and decoding the final speech output. 
In our experiments, we adopted DAC \cite{kumar2023high} as the neural speech codec. 
The coding results of clean speech from DAC {serve} as the upper bound for the enhancement quality achievable by UDSE. 
Theoretically, UDSE can also work with other RVQ-based neural speech codecs that offer comparable coding quality. 
To verify the generalization ability of UDSE across different neural speech codecs, we replaced the DAC with a more lightweight RVQ-based MDCTCodec \cite{jiang2024mdctcodec} in UDSE (denoted by UDSE w/ MDCTCodec). 
The results in Table \ref{tab:a} showed no significant difference between UDSE and UDSE w/ MDCTCodec ($p=0.64$), indicating that replacing DAC with MDCTCodec did not obviously affect the performance of UDSE. 
This confirms that UDSE exhibits strong codec generalization to different RVQ-based codecs, laying a solid foundation for future efforts to reduce the complexity of UDSE by exploring more lightweight codec alternatives.

\section{Conclusion}
\label{sec:con}

This paper proposed UDSE, a novel classification-based discrete-domain general universal SE model. 
Unlike conventional regression-based continuous-domain SE models that predict continuous features/waveforms, UDSE formulates the SE task as a classification problem over clean acoustic tokens quantized by a neural speech codec. 
The UDSE advances discrete-domain SE by introducing a self-contained framework that does not require any textual cues, semantic labels, or support from LLMs. 
The UDSE begins with a randomly initialized token sequence, extracts global feature conditions from degraded speech, and sequentially predicts clean tokens quantized by each VQ of a neural speech codec following the RVQ rule, where each prediction depends on the previous ones.
The final clean speech is reconstructed by decoding all predicted tokens using the decoder of the neural speech codec.
During training, UDSE is optimized using teacher forcing strategy with a cross-entropy classification loss. 
Both objective and subjective experimental results demonstrate that the proposed UDSE can effectively enhance speech affected by various single and combined distortions, confirming its strong universality and practical applicability. 
{In future work, we will further optimize the model architecture and design to reduce computational complexity and latency, while continuing to explore enhancement performance under a wider range of distortion scenarios.
In addition, we will explore applying the UDSE token prediction framework to other speech tasks, such as zero-shot TTS with LLMs, to extend its applicability.
}

\bibliographystyle{IEEEtran}
\bibliography{mybib}
\end{document}